# Fast Multi-Scale Detection of Relevant Communities

Erwan Le Martelot

Department of Computing

Imperial College London

London SW7 2AZ, United Kingdom

e.le-martelot@imperial.ac.uk

Chris Hankin

Department of Computing

Imperial College London

London SW7 2AZ, United Kingdom

c.hankin@imperial.ac.uk

October 30, 2018

**Abstract**

Nowadays, networks are almost ubiquitous. In the past decade, community detection received an increasing interest as a way to uncover the structure of networks by grouping nodes into communities more densely connected internally than externally. Yet most of the effective methods available do not consider the potential levels of organisation, or scales, a network may encompass and are therefore limited. In this paper we present a method compatible with global and local criteria that enables fast multi-scale community detection. The method is derived in two algorithms, one for each type of criterion, and implemented with 6 known criteria. Uncovering communities at various scales is a computationally expensive task. Therefore this work puts a strong emphasis on the reduction of computational complexity. Some heuristics are introduced for speed-up purposes. Experiments demonstrate the efficiency and accuracy of our method with respect to each algorithm and criterion by testing them against large generated multi-scale networks. This study also offers a comparison between criteria and between the global and local approaches.

## 1 Introduction

Whether studying the structure of the Internet, telephone networks, power grids, transportation networks, protein interactions or social networks, we are most likely studying graphs, or networks [14]. Consequently network science became a wide-reaching field and advances in this domain can contribute to advances in many others. Within network science, the field of community detection attracted a lot of interest in the past decade considering community structures as important features of real-world networks [4]. Commonly, community detection refers to finding groups of nodes more densely connected internally than externally. This concept can be approached from a global perspective (i.e. considering networks as a whole) or from a local perspective (i.e. exploring network areas with no global visibility). Boundaries between communities can be considered sharp or overlapping. Tunings can be used to bias the detection towards clusters of various sizes. Community detection can therefore be approached in several ways. This resulted in the creation of various methods to address the problem [4]. In general, community detection methods use a criterion to rank communities and an optimisation algorithm to process the data. The algorithms often rely on heuristics in order to process the data in a reasonable amount of time. Indeed the partitioning of a graph is an NP-hard task [4] and datasets in real-world problems are often large. Therefore a significant emphasis is put on producing algorithms with a low complexity.

Also networks often have several levels of organisation [18], leading to different relevant communities at various scales (or resolutions). Hence identifying the right scales becomes an additional part of the problem. Accurate community detection in a network thus implies uncovering communities at identified scales of relevance. While many approaches have been presented for community detection few address this multi-scale issue. Two well known criteria dealing with scales are two variations of modularity proposed by Arenas et al. [1] and Reichardt and Bornholdt [15]. Recently, stability optimisation [10] based on a measure called *stability* [3] was investigated. It uses random walks of various length on a network to enable multi-scale analysis. Another method introduced in [16, 17] considers intra-cluster connectivity and provides a tuning parameter to aim at communities of various sizes. These four approaches are global approaches. However some local community detection methods with a resolution parameter have also been presented [9, 6].

Thus several criteria have been suggested to detect communities given a scale parameter. A first multi-resolution method where scales of relevance are found by minimising the difference between the results of several runs per scale was suggested in [16]. Yet the method relies on replications and does not reuse computed information across scales. These computational redundancies thus limit its efficiency. Therefore no method exists to accurately and efficiently uncover communities across scales and identify the communities and scales of relevance. Such a method would enable analysts to quickly extract from unknown data the significant communities and scales of interest. This paper addresses this issue and presents a method that enables fast detection of communities across scales. The method is derived in two algorithms, one designed for global criteria, the other one for local criteria. The paper also introduces heuristics designed for the efficient optimisation of some criteria.

The following section reviews the relevant contributions found in the literature in the field of community detection. Then our method, algorithms and heuristics are presented followed by experiments. Experiments are performed on large networks and assess our method both from an accuracy and a scalability point of view. The paper then concludes on the results of this work and its implications for future work.

## 2 Background

In the recent years several multi-scale criteria were introduced. The two best known are probably variations of Newman's modularity [13] by Reichardt and Bornholdt [15] and Arenas et al. [1].

Modularity is the sum of the difference between the fraction of links within a partition linking to this very partition minus the expected value of the fraction of links doing so if edges were randomly placed, as given in equation (1) where $A$ is the adjacency matrix, $m$ the number of edges (or total strength for a weighted network), $d_i$ the degree (or strength) of node $i$, and the $\delta(i,j)$ function returns one if nodes $i$ and $j$ belong to the same community, zero otherwise.

$$Q_M = \frac{1}{2m} \sum_{i,j} (A_{ij} - \frac{d_i d_j}{2m}) \delta(i,j) \tag{1}$$

In [15], the authors modify modularity by using a scalar parameter $\gamma$ in front of the null term (the fraction of edges connecting vertices of a same community in a random graph) turning equation (1) into

$$Q_{M_\gamma} = \frac{1}{2m} \sum_{i,j} (A_{ij} - \gamma \frac{d_i d_j}{2m}) \delta(i,j) \tag{2}$$

where $\gamma$ can be varied to alter the importance given to the null term (modularity optimisation is found for $\gamma = 1$). In [1], modularity optimisation is performed on a network where each node's strength has been reinforced with self loops. Considering the adjacency matrix $A$, modularity optimisation is performed on $A + rI$ where $I$ is the identity matrix and $r$ is a scalar:

$$Q_{M_r} = Q_M(A + rI) \qquad (3)$$

Varying the value of $r$ enables the detection of communities at various coarseness levels (modularity optimisation is found for $r = 0$).

Recently, a new partition quality measure called *stability* was introduced in [3] and its use as an optimisation criterion was investigated in [10]. The stability of a graph considers the graph as a Markov chain where each node represents a state and each edge a possible state transition. Let $d$ be the degree vector giving for each node its degree (or strength for a weighted network) and $D = diag(d)$ the corresponding diagonal matrix. The stability of a graph considers the graph as a Markov chain where each node represents a state and each edge a possible state transition. The transition probabilities between states is given by the $n \times n$ stochastic matrix $M = D^{-1}A$. The transition probabilities for a random walk of length $t$ ($t$ is the Markov time) are then given by $M^t$. Then $A_t = D \cdot M^t$ gives the edges of the network representing a random walk of length $t$ on the original network. Following the method from [10] stability for a walk of length $t$ can be expressed similarly to the modularity expression from equation (1) as

$$Q_{M_t} = \frac{1}{2m} \sum_{i,j} (A_{t_{ij}} - \frac{d_i d_j}{2m}) \delta(i,j) \qquad (4)$$

where the adjacency matrix is the matrix $A_t$. Additionally, decimal values of $t$ can be computed between two successive integer values of $t$ by using the linear interpolation:

$$A_t = (\lceil t \rceil - t) \cdot A(\lfloor t \rfloor) + (t - \lfloor t \rfloor) \cdot A(\lceil t \rceil) \qquad (5)$$

where $\lceil t \rceil$ returns the smallest integer greater than $t$ and $\lfloor t \rfloor$ returns the greatest integer smaller than $t$. This is particularly useful to investigate time values between 0 and 1 as studies from [3, 10] show that the use of Markov time within this interval enables detecting fine partitions.

Another multi-scale method, not relying on modularity, was introduced in [17]. The model uses no null factor and is therefore not subject to the resolution limit found in modularity. The quality of a partition is expressed as

$$Q_H(\gamma) = -\frac{1}{2} \sum_{i \neq j} (A_{ij} - \gamma J_{ij}) \delta(i,j) \qquad (6)$$

where $A$ is the adjacency matrix and $J_{ij} = 1 - A_{ij}$ (i.e. $J$ is the complement of the adjacency matrix $A$) and $\gamma$ is the resolution parameter. The models therefore considers the amount of connections within communities less the connections missing to get fully connected communities. $\gamma$ varies the importance of the missing connections. A small $\gamma$ value will favour large communities while a large $\gamma$ value will favour dense ones. (Note that this model considers only local information at the community level. It is optimised in a global manner though.)

The first two methods from Reichardt and Bornholdt [15] and Arenas et al. [1] share somewhat the idea that modifying the impact of some factors within the modularity equation (e.g. the null factor, nodes weight with self-loops) can offer a multi-scale approach to community detection using modularity. However neither of these two criteria enable an in-depth exploration of the network. The former two criteria remain based on a one step random walk analysis of the network with modifications of its structure or of the null factor. The method from Ronhovde and Nussinov [17] rewards compactness. In contrast, stability optimisation [10] enables random walks of variable length thus exploiting the actual structure of the network similarly to an information flow through. The correlation with Markov chains also provides mathematical foundations that give the scale parameter an actual meaning.

The three criteria derived from modularity also have in common that they have all been optimised in the literature using algorithms derived from Newman's fast algorithm [11] which is a pairwise greedy aggregation method. The criterion from Ronhovde and Nussinov used a different greedy approach with several runs to select the best one out of them. In terms of algorithms one of the best known efficient algorithms for community detection is the Louvain method [2] which optimises modularity and can deal with large graphs. However this algorithm does not deal with multi-scale community detection.

Other approaches based on local criteria can also deal with multi-scale community detection. In [9] the authors introduce the fitness of a community $c$ as

$$f_c = \frac{k_{in}^c}{(k_{in}^c + k_{out}^c)^\alpha} \tag{7}$$

and then test whether a node $i$ should join a community $c$ by computing the fitness of $i$ with respect to $c$ as

$$f_c^i = f_{c+i} - f_{c-i} \tag{8}$$

The parameter $\alpha$ sets the scale of the method. Large values of $\alpha$ lead to small communities while small values lead to large ones.

Another approach presented in [6] uses the structural similarity between nodes given by

$$s(i,j) = \frac{\sum_{k \in \Gamma(i) \cap \Gamma(j)} w(i,k) \cdot w(k,j)}{\sqrt{\sum_{k \in \Gamma(i)} w(i,k)^2} \cdot \sqrt{\sum_{k \in \Gamma(j)} w(k,j)^2}} \tag{9}$$

where $\Gamma(i)$ is the neighbourhood of node $i$ and $w(i,j)$ the weight between nodes $i$ and $j$ (equivalent to the adjacency matrix notation $A_{ij}$). The tightness $T$ of a community is then calculated as

$$T_c = \frac{S_{in}^c}{S_{in}^c + S_{out}^c} \tag{10}$$

where $S_{in}^c = \sum_{i \in c, j \in c} s(i,j)$ is the internal similarity of the community $c$ and $S_{out}^c = \sum_{i \in c, j \notin c} s(i,j)$ is its external similarity. The test regarding whether or not a node $i$ should join a community $c$ is similar to equation (8). Following the method from the authors, the criterion to optimise is the tightness gain given by

$$\tau_c^\alpha(i) = \frac{S_{out}^c}{S_{in}^c} - \frac{\alpha S_{out}^c - S_{in}^c}{2S_{in}^c} \tag{11}$$

where $\alpha$ is the tuning parameter.

The two latter methods optimise communities using a local criterion. Compared to the global criteria presented above such approaches enable to grow communities from nodes with knowledge of only their neighbourhood. The starting node may therefore play a significant role in the final shape of each community. Also since each community growth is an independent process, the resulting communities can share nodes and thus be overlapping.

The next section presents our method, algorithms and heuristics for fast multi-scale community detection using all the aforementioned criteria. We then provide an experiment section comparing the performance and accuracy of the algorithm with the various criteria.

# 3 Method

We have seen in the previous section that several criteria with a resolution parameter have been suggested. While the main aim of our method is speed efficiency it is also desirable to keep the method criterion neutral (i.e. enable the usage of any criterion) as much as possible. The final aim is also to exploit the results to identify the relevant communities and scales.

To deal with the speed aspect we chose to use a greedy approach that exploits all the available information (i.e. input data and information computed as the algorithm runs). To do so we consider that the outcome of the algorithm for a specific parameter value is valuable information that can be exploited for further parameter values. More specifically the result for parameter value $p$ could help uncover the result for the following parameter $p + \delta p$.

Our method will be based on an aggregation process that builds larger and larger communities as parameters are given in order of increasing scale. Therefore the input parameter list much be such that $\forall (i,j) \in \mathbb{N}^2 : i < j \Rightarrow scale(p_i) < scale(p_j)$. For each parameter $p_j$ following $p_i$ the algorithm will start its computation based on the outcome for $p_i$ instead of starting from scratch. Variations between successive sets of communities may sometimes be small or large. Therefore our algorithm needs to allow for both small and larger changes to occur in an efficient way. This can be done by using a two phase approach where one phase performs subtle changes at the node level and the second phase performs coarser operations at the community level. These phases can alternate until no further refinement is possible.

Considering a global criterion approach with crisp boundaries between communities, subtle changes can be made by shifting nodes from their current community to others in the neighbourhood. Larger and coarser changes can be made by merging communities should this operation improve the criterion value. In case of significant changes due to successive scales this (second) phase will provide a fast algorithm progression. Should some nodes then be placed in a non-optimum location, going back to the node shifting phase can correct this. The first phase can remind us of the Kernighan-Lin (KL) algorithm [7] that aims at minimising the total edge weights across clusters in a network by repeatedly swapping nodes belonging to different clusters that yields a maximum weight cut reduction. The KL algorithm has been adapted as a refinement process for community partitions by Newman [12]. In Newman's version each node on the edge between two communities is put in the other community to test if the move would result in a modularity increase. This idea is also present in the Louvain method [2]. This concept is reused here by taking as input a previously computed set of communities and moving nodes from their community to neighbour communities if the move results in an increase in the value of the criterion being optimised. The merging phase is somehow similar to Newman's fast modularity optimisation method [11] where communities are successively merged in a hierarchical manner. However here the merging only takes place as long as it results in an increase of the criterion value.

Regarding local criteria, the first phase of subtle changes can be performed using a growth function (see further below) that expands communities until the local criterion can no longer be improved. This can be extended to growing communities of any size. The larger change phase can then involve merging communities that overlap significantly, thus reducing the amount of communities while maintaining their integrity.

Therefore the method alternates between small and coarse changes until no further optimisation is possible. The best set of communities for the current parameter has then been found and the next scale parameter is taken, using the current communities as starting point. Assuming that most of the current community distribution may remain unchanged, the amount of changes between two successive scales can be minimal (e.g. a few moves), thus significantly reducing the computation required for each scale. (Note also that the method can be used for mono-scale community detection by being given a unique parameter. Modularity can be optimised for instance by giving 1 to the implementations using Reichardt and Bornholdt's criterion or stability optimisation.)

Based on the described method we propose below the pseudo-code for two derived algorithms, one for global criteria, the other for local criteria.

## 3.1 Algorithm for Global Criteria

The pseudo-code of the global criteria based algorithm is given in Algorithm 1.

---
**Algorithm 1** Fast multi-scale community detection algorithm for global criteria.
---
1: Initialise current community partition with a node per community: $com$ = list of all nodes
2: **for all** scale parameters $p$ **do**
3:     Compute initial $Q$ value given $p$: $Q = computeQ(com, p)$
4:     **while** changes can be made **do**
5:         **while** nodes can be moved **do**
6:             $nlist$ = list of all nodes
7:             **while** $nlist$ is not empty **do**
8:                 $n$ = pick a random node in $nlist$
9:                 $ncom$ = neighbour communities of $n$
10:                 $best\_\Delta Q = 0$
11:                 **for all** communities $nc$ in $ncom$ **do**
12:                     Compute the $\Delta Q$ that moving $n$ into $nc$ would produce
13:                     **if** $\Delta Q > best\_\Delta Q$ and move does not break a community **then**
14:                         $best\_\Delta Q = \Delta Q$
15:                         $best\_c = nc$
16:                     **end if**
17:                 **end for**
18:                 **if** $best\_\Delta Q > 0$ **then**
19:                     Update $com$: move node $n$ to community $best\_c$
20:                     Update total value of $Q$: $Q = Q + best\_\Delta Q$
21:                 **end if**
22:             **end while**
23:         **end while**
24:         **while** clusters can be merged **do**
25:             $clist$ = list of all current communities
26:             **while** $clist$ is not empty **do**
27:                 $c$ = pick a random community in $clist$
28:                 $ncom$ = neighbour communities of $c$
29:                 $best\_\Delta Q = 0$
30:                 **for all** communities $nc$ in $ncom$ **do**
31:                     Compute the $\Delta Q$ that moving $n$ into $nc$ would produce
32:                     **if** $\Delta Q > best\_\Delta Q$ **then**
33:                         $best\_\Delta Q = \Delta Q$
34:                         $best\_c = nc$
35:                     **end if**
36:                 **end for**
37:                 **if** $best\_\Delta Q > 0$ **then**
38:                     Update $com$: merge communities $c$ and $best\_c$
39:                     Update total value of $Q$: $Q = Q + best\_\Delta Q$
40:                 **end if**
41:             **end while**
42:         **end while**
43:     **end while**
44:     Store $com$ and $Q$ for $p$
45: **end for**
46: **return** Community sets and associated $Qs$
---

This algorithm can be used to optimise the criteria from [15, 1, 17, 10]. In is important to note that in the first phase, when moving a node from a community to another it may be necessary to check that the original community remains a connected component after the change (line 13). If the node being removed from the community was the only connection between two subgroups of nodes within this community, then the node should not be removed as otherwise the community becomes two independent components which violates the concept of community. This degenerate case may sometimes not be allowed by the criterion itself (i.e. $\Delta Q$ cannot be positive in this situation) in which case no additional test is needed. However criteria are usually designed to rank valid sets of communities and not to detect community structure errors. Therefore no such guarantee on $\Delta Q$ exists. To guarantee the integrity of the communities we added a test using a breadth-first search method that starts from any node in a community and checks that all the nodes can be reached. Any candidate move (with $\Delta Q > 0$) not passing this test will be discarded.

The overall complexity of the algorithm is not straightforward to establish. However, the first phase iterates through nodes and considers their neighbourhood, hence the edges. Therefore this phase is in $\mathcal{O}(m)$ (with $m$ the number of edges or total weight of the network). The second phase considers communities and attempts to merge neighbour communities. Again this will be determined based on the edges and we can say that the phase is in $\mathcal{O}(m)$. The amount of times these two phases will be repeated for each parameter value varies but is expected to be low (e.g. once, twice most of the time). Hence for each parameter the complexity is in $\mathcal{O}(m)$. As previously discussed, each new set of communities is computed from the previous one thus reducing the amount of additional computation. For $p$ parameters the overall complexity will remain $\mathcal{O}(m)$.

## 3.2 Algorithm for Local Criteria

The pseudo-code of the local criteria based algorithm is given in Algorithm 2. The main difference between the global and local algorithm is in the use of the criterion. The local algorithm only uses the criterion in the first phase in the function that grows communities, whether from a single node (seed) or from an existing community. This function is thus criterion dependent and its implementation is not fixed. In our work, for the criterion from Huang et al. we followed the growth method from the authors [6]. However for Lancichinetti et al.'s criterion we wrote a new growth method (see implementation details in Algorithm 4) as the one described in [9] has a high complexity. The idea for growing communities using a local criterion is usually to start from an initial node (called a seed) or an existing community and then grow the community by successively adding neighbour nodes that improve the criterion value until no node can be added. This is similar to the first phase of Algorithm 1 but here when a node is added to a community it is not taken from another community, thus enabling overlapping communities. Note however that the possibility of getting overlapping communities can be turned off by not allowing to add to a community a node that is already member of another community.

Local criteria are designed to consider the addition or removal of nodes to a community in order to perform a growth process. They are not designed to assess larger operations such as the merging of several communities. For instance, considering equation (7) for Lancichinetti et al.'s criterion and equation (10) for Huang et al.'s criterion show that the set of communities with one community encompassing all the nodes maximises both equations. Therefore the second phase does not rely on the local criterion but instead consists in merging communities if they overlap significantly (as opposed to improving a global criterion in the global criterion algorithm). Indeed as communities grow independently from one another in the first phase, some may overlap. The overlap ratio for merging is controlled by a threshold $\eta$. Two communities $C_1$ and $C_2$ are merged if $\frac{|C_1 \cap C_2|}{|C_2|} \geq \eta$ or $\frac{|C_1 \cap C_2|}{|C_1|} \geq \eta$. ($|C|$ refers to the cardinality of $C$.) By default we set $\eta = 0.5$

**Algorithm 2** Fast multi-scale community detection algorithm for local criteria.

```
1:  for all scale parameters p do
2:      if is p is the first parameter then
3:          Initialise all nodes as potential seeds: slist = list of all seeds
4:          while slist is not empty do
5:              Initialise new community c with a seed n and remove n from slist
6:              Grow c from n according to the criterion tuned by p
7:              Remove from slist the seed n and the nodes from c
8:          end while
9:      else
10:         for all communities c do
11:             Grow c according to the criterion tuned by p
12:             if c changed then
13:                 Add c to the set C_M of communities to check for merging
14:                 if c encompasses another community c_2 not grown yet in this loop then
15:                     Remove c_2
16:                 end if
17:             end if
18:         end for
19:     end if
20:     while C_M is not empty do
21:         Extract first community c_1 from C_M
22:         for all other existing communities c_2 do
23:             if c_1 and c_2 have a shared nodes ratio ≥ η then
24:                 Merge c_1 and c_2 into c_1
25:                 Add c_1 to the set C_M
26:             end if
27:         end for
28:     end while
29:     Store com and Q for p
30: end for
31: return Community sets and associated Qs
```

so two communities are merged if they overlap on half of the nodes or more of at least one of them. Note that when dealing with weighted networks the cardinality can be replaced by the sum of edge weights within the communities. Note also that in line 14 the algorithm checks if the last grown community encompasses some of the remaining communities in the loop. If so these smaller communities are removed as growing them will most likely result in a very similar community to the one just grown and thus waste computing time.

Here, the global $Q$ value for the set of communities is computed as the average of the Q values (local criteria) over all communities.

Regarding complexity, the larger the network the more communities can potentially be grown. The complexity of the growth function depends on its implementation and is discussed in the next section. The complexity of the first phase is therefore difficult to analyse but depends on the complexity of the growth function and the size of the network. The complexity of the community merging phase can however be estimated. Lines 20 and 22 form two nested loops iterating through the number of communities $n_c$. Computing the number of shared nodes depends on the size of the communities. If we consider the average community size $k$ over all communities the complexity is $\mathcal{O}(n_c^2 \cdot k)$. As communities grow (i.e. as $k$ grows), $n_c$ decreases since communities are merged when their node overlapping ratio reaches $\eta$. This leads to the constraint

$$k < \eta \cdot k + \frac{n - \eta \cdot k}{n_c} \qquad (12)$$

By turning equation (12) into an equality we can compute an upper bound for $n_c$ to maximise $n_c^2 \cdot k$. It also has to follow the constraints $n_c \cdot k \geq n$ and $\eta \cdot k \leq n$. With $\eta = 0.5$, the expression $n_c^2 \cdot k$ is maximised for $n_c = n-1$ and $k = 2$ and minimised for $n_c = 2$ and $k = \frac{2}{3}n$. The worst case complexity is therefore $\mathcal{O}(n^2)$ while the best case complexity is $\mathcal{O}(n)$. The merging phase tends to take more time for scales with small communities than for scales with large communities. It is noteworthy that this second phase can be removed if communities are not allowed to overlap. Therefore, enabling overlapping communities adds a significant complexity overhead. On the other hand, preventing the growth of a community based on other communities may reduce the accuracy performance. A system of node swapping like for the first phase of the global criteria algorithm would then be needed. This would also require to find another merging criterion as the one based on overlapping nodes could no longer be used. As the global criteria algorithm already addresses the fast detection of non-overlapping communities we favoured allowing the communities to overlap in the local criteria algorithm, to the cost of efficiency.

### 3.3 Implementation Choices and Heuristics

This section details some implementation choices that make the aforementioned complexities reachable. The algorithm makes extensive use of the neighbourhood of nodes, hence networks are represented using an adjacency list, giving for each node a list of pairs containing the target node and the edge value, sorted by node index value. This also enables iterating in $\mathcal{O}(m)$ through the edges. The communities are stored as lists of nodes they respectively contain.

#### 3.3.1 Global Criteria

In the global criteria implementation an additional array provides the community membership for each node node. The sum of internal weights (sum of edges between members) and the total weight (sum of edges from a member to anywhere) of communities are kept in arrays and updated at each change. The only information that is computed for each potential move is the variation $\Delta Q$ of the criterion. Considering the modularity expression given in equation (1), we can derive that moving a node $i$ from a community $c_i$ to another community $c_*$ produces

$$\Delta Q_{i,c_i,c_*} = \frac{1}{2m}\left(-(\sum_{k \in c_i} A_{ik} - A_{ii}) + 2 \cdot \frac{d_i \cdot (W_{c_i} - d_i)}{2m} + \sum_{k \in c_*} A_{ik} - 2 \cdot \frac{d_i \cdot W_{c_*}}{2m}\right) \quad (13)$$

where $W_c$ is the total weight of community $c$. The first term in the parenthesis expresses that moving node $i$ from its community $c_i$ takes its contribution from the internal weight (note that the self-loops $A_{ii}$ are not removed as they contribute to whatever community $i$ is part of). The second term gives back to $\Delta Q$ the contribution to the null term that was removed when adding $i$ to $c_i$. The third term adds the contribution from $i$ to the community $c_*$. The last term removes the associated null factor contribution. Regarding phase 2, the fast modularity optimisation equation from [11] is used. For the criteria from Reichardt and Bornholdt, Arenas et al. and stability optimisation, the equations are modified accordingly to take into account the tuning parameter. Equations with the same principle are derived for Ronhovde and Nussinov's criterion.

**Stability optimisation:** Implementing stability optimisation efficiently is difficult when we need to compute the transition probabilities between nodes for walks of variable length $t > 1$ ($M^t$ in the matrix notation). In the best case scenario we can compute $M^{t_1+t_2} = M^{t_1} \cdot M^{t_2}$ if we already know $M^{t_1}$ and $M^{t_2}$ (e.g. compute $M^3$ knowing already $M^2$ and $M$). We then only need to perform one matrix multiplication. (We also implemented a procedure computing a decomposition minimising the number of $M^t$ computations.) By multiplying a matrix using Strassen's algorithm [19], computing $M^{t_1+t_2}$ has a complexity of about $\mathcal{O}(n^{2.807})$ which is good for dense networks but can be improved for sparse networks. To do so we dropped the matrix representation and used the adjacency list representation for the computation of the new edges. Assuming we know the edges for the walks of lengths $t_1$ and $t_2$, the resulting edges can be computed as given in Algorithm 3 in lines 1-10. As the edges computation for each nodes is independent this algorithm can easily be parallelised by dividing the number of nodes into groups and give each group to a separate thread.

Regarding the complexity of the operation, let's assume this algorithm will most often be used to compute the edges for the current length incremented by 1. (This assumption allows a clearer analysis.) Therefore the adjacency list $al_t$ will be computed from edges at length $t-1$ and from the original edges of the network. Let $\bar{d}$ be the average node degree in the network. For a walk of length 2 each edge is combined with an adjacent edge to create the new edges. Assuming the network is sparse this yields approximately $\bar{d}^2$ edges. By induction we get that for a walk of time $t$ the network's number of edges is proportional to $\bar{d}^t$ with an upper bound at $n-1$ which is the maximum degree yielding a fully connected network. Therefore the complexity of the computation of $al_t$ from $al_{t-1}$ and $al_1$ is $\mathcal{O}(n \cdot \bar{d}^t)$. For values of $t$ large enough to yield a fully connected network this tends towards $\mathcal{O}(n^2 \cdot \bar{d})$ which for a sparse network (i.e. $\bar{d}$ is small) is $\mathcal{O}(n^2)$. However for dense networks the complexity becomes $\mathcal{O}(n^3)$ and then using Strassen's algorithm on matrices is more efficient. The space complexity is also $\mathcal{O}(n \cdot \bar{d}^t)$ and tends towards $\mathcal{O}(n^2)$ as $t$ increases.

To further speed up the edge computation process we introduce an edge threshold $\tau$. Edges with a value below $\tau$ are not added to the adjacency list $al_{t_1+t_2}$. This heuristic (given in Algorithm 3 in lines 11-15) reduces the increase in complexity and memory usage that occurs as $t$ increases. This comes to the cost of a potential accuracy loss. As edge values reflect data information there is no ad-hoc value for this threshold and its setup is left to the user. However some insight can be gained by considering its meaning with respect to the network structure. For instance if the initial edges have a value of 1 and the average node degree is $\bar{d} = 10$, then at $t = 2$ the lowest edges will have a value of 0.1, at $t = 3$ a value of 0.01, etc. Thus filtering edge values below say 0.01 discards the weakest edges that can be created within three steps. At each increase of $t$ by 1 the number of edges increases by a factor of up to 10 which for large graphs is a significant increase in complexity and memory resource. The threshold $\tau$ enables to reduce this increase.

**Algorithm 3** Computation of adjacency list $al_{t_1+t_2}$ for a random walk of length $t_1 + t_2$ given the adjacency lists $al_{t_1}$ and $al_{t_2}$.

```
 1: for all nodes n do
 2:     for all edges starting from n in the adjacency list al_{t_1} do
 3:         n_1 = current neighbour node of n in al_{t_1}
 4:         Compute transition probability between n and n_1: v_1 = al_{t_1}(n, n_1)/degree(n)
 5:         for all edges starting from n_1 in the adjacency list al_{t_2} do
 6:             n_2 = current neighbour node of n in al_{t_2}
 7:             Compute transition probability between n_1 and n_2: v_2 = al_{t_2}(n_1, n_2)/degree(n_1)
 8:             al_{t_1+t_2}(n, n_2) = al_{t_1+t_2}(n, n_2) + degree(n) · v_1 · v_2
 9:         end for
10:     end for
11:     for all edges going from n to k in the adjacency list al_{t_1+t_2} do
12:         if al_{t_1+t_2}(n, k) < τ then
13:             Remove al_{t_1+t_2}(n, k)
14:         end if
15:     end for
16: end for
```

### 3.3.2 Local Criteria

For the criterion from Lancichinetti et al. the sum of internal weights and the total weight of each communities are kept in arrays and updated at each change. For the criterion from Huang et al. the weights are replaced by the similarities between vertices as given in equation (9). These similarities are computed for all pairs once for all nodes before the execution of the algorithm in $\mathcal{O}(n \cdot \bar{d}^2) = \mathcal{O}(m \cdot \bar{d})$. For each community being grown a set of nodes (represented as a binary search tree) sorted by the node index maintains the list of neighbours of the community that are potential candidates for joining. The nodes of each community remain sorted by node index, allowing operations like merging maintaining the sorted property or getting the overlapping nodes to be performed in linear time. Communities are also converted to a bit vector during growing and merging operations as these require many lookup, insert and delete operations. Maintaining the bit vector structure allows to perform these operations in $\mathcal{O}(1)$ thus speeding up significantly the process to the cost of some extra memory. The growth process uses the bit vector representation only and then converts the community back to a sorted list in $\mathcal{O}(n.log(n))$. The merging process only requires lookup operations before merging. It can hence make use of both representations at the same time to get fast lookups and keep the node lists sorted when merging in $\mathcal{O}(n)$, thus avoiding another conversion from the bit vector to a sorted list in $\mathcal{O}(n.log(n))$.

Regarding the method from Lancichinetti et al., the growth function suggested by the authors (see [9]) works as follows. Given a community to grow, while nodes can potentially be added to it, the function considers all neighbours of the community and picks the one that increases most the criterion value. In no node qualifies, the growth stops. If a node qualifies it is added to the community and for all the nodes in the community the function checks whether they still contribute positively to the criterion value, failing which they are removed. The amount of neighbour nodes as well as the number of nodes in the community grows in $\mathcal{O}(n)$. As this function contains 3 nested loops we can get to $\mathcal{O}(n^2 \cdot m)$ which is high for large networks. (With the neighbour list to maintain it adds up to $\mathcal{O}(n^2 \cdot m \cdot \bar{d})$.) To reduce this we can check whether nodes should stay or not in the community only after all nodes have been added. We now only have twice 2 nested loops bringing the complexity down to $\mathcal{O}(n \cdot m \cdot \bar{d})$. To further speed up the process we can simplify the node checking by checking all nodes once only, instead of rechecking from the beginning after a node has been removed, and repeat this process a maximum of $k$ times. This bring the checking loop complexity to $\mathcal{O}(k \cdot m \cdot \bar{d})$. By default our implementation does not constrain $k$ (i.e. $k = \infty$) as in practice the number of times the check is performed is low giving an expected complexity of $\mathcal{O}(m \cdot \bar{d})$. Adding nodes however remains costly as at each pass all neighbours are considered in order to find the best. To speed up this part we store the community neighbours in a max priority queue using the factor $\frac{2.d_{in}}{(d_{in}+d_{out})^\alpha}$ to rank nodes, where $d_{in}$ is the sum of edge weights from a node to a community and $d_{out}$ the remaining edge weights of the node. This heuristic enables taking the neighbour nodes in an order of overall decreasing impact on the criterion. This allows to perform a single pass only through the neighbours set in expected $\mathcal{O}(m \cdot \bar{d})$ with a minimal loss of performance. (Note that a priority queue is also used in Huang et al.'s method where their similarity criterion fits perfectly as a ranking order. See [6] for details.) The overall expected complexity of our growth function is thus $\mathcal{O}(m \cdot \bar{d})$. Note however that the set of neighbour nodes is a subset of the node set, hence it contains at most $n - 1$ nodes and most often contains less. Our modified function is given in Algorithm 4.

In line 14, Algorithm 2 checks whether community $c$ encompasses community $c_2$. This function iterates through the (sorted) nodes lists of the communities and returns false as soon as one node in $c_2$ is not matched in $c$. It returns true if all nodes of $c_2$ have ben matched.

**Algorithm 4** Fast method to grow a community $c$ using the criterion from [9].
1: Create neighbour nodes max priority queue using factor $\frac{2.d_{in}}{(d_{in}+d_{out})^\alpha}$
2: **while** priority queue is not empty **do**
3:     Pick first node $n$
4:     **if** $n$ improves $Q_c$ **then**
5:         Add $n$ to $c$
6:         Update or add in priority queue neighbours of $n$ not in $c$
7:     **end if**
8: **end while**
9: **if** a node has been added **then**
10:     **while** Number of iterations $< k$ **do**
11:         **for all** nodes $n$ in $c$ **do**
12:             Recompute $Q_{c\backslash n}$
13:             **if** $Q_{c\backslash n} > Q_c$ **then**
14:                 $n$ is removed from $c$
15:             **end if**
16:         **end for**
17:         Exit *while* loop if no node could be removed
18:     **end while**
19: **end if**

# 4 Experiments

The following section presents sets of experiments that have been performed to assess our method. For both the global and local algorithm and for each criterion a dedicated implementation was coded in C++ as well as in Matlab. In the following experiments we use the C++ implementations[1]. All experiments were run under MacOS X 10.7.3 on a desktop computer iMac 3.06GHz Intel Core i3 with 4GB of RAM.[2]

The aim of our method is to provide an efficient tool for the analysis of unknown potentially large networks. The algorithm must be accurate but also efficient to provide in a short amount of time some community sets that can potentially be further analysed using computational tools, visualisation methods, or any other relevant method. Therefore both accuracy and efficiency will be assessed.

In order to test the algorithm's performance and perform a comparative analysis of the criteria we used the benchmark from Lancichinetti and Fortunato [8] that was designed to provide networks with communities at both micro and macro scales.

Regarding the scale parameters, in all experiments we use a logarithmic sampling of the possible values within the interval of relevance to each criterion. The scale sampling is given by

$$Values = A \cdot \frac{1 - log([1:1:X])}{log(X)}$$

where X is the number of values we want in the sample, $[1:1:X]$ the vector of values between 1 and $X$ incremented by 1 between each value. The formula returns a vector of $X$ sample values within the interval $[0, A]$ with values around 0 close to each other and then progressively spreading out towards $A$. Each criterion has a range of scales of relevance which may vary between different networks. (However they tend to remain fairly similar across networks as experiments will show.)

---

[1]The code developed for this work is available for download from http://www.elemartelot.org. All algorithms are available as well as a flexible testing framework in which any other algorithm can be added. See documentation on-line.

[2]In the current version, the code is not multi-threaded (except for Algorithm 3 which computes the networks in stability optimisation). Performance could be improved further with multi-threading and a greater amount of RAM on the test machine. For the purpose of this work however the current implementation is very efficient on a desktop machine and its code is compatible with most systems, making the tool accessible for everybody.

The information change between community sets is measured using the normalised mutual information (NMI) [5] when communities are crisp, and using the alternative definition from [9] when communities are overlapping. To analyse how much change there is between successive community sets we measure the NMI averaged over $p$ successive scales. We use $p = 3$ and $p = 5$ in our experiments. A short range reveals a potentially short consistency between community sets while a longer range reveals longer consistencies. The longer the consistency the more robust to scale variation a community set is, and the more confidence we can have in the relevance of the set.

## 4.1 Accuracy Testing

The network generator can be tuned to generate networks with various statistics by varying the number of nodes, the average node degree, the maximum node degree, the ratio of internal edges respectfully in the micro and in the macro communities, and their minimum and maximum sizes. The generator returns the intended community sets. This enables non-biased accuracy assessment by direct comparison between the uncovered communities and the intended ones.

The aim of multi-scale community analysis is to uncover communities at the relevant scales. This does not necessarily imply uncovering several community sets at several scales. However the use of a network with micro and macro scales guarantees communities can be found at different scales, thus making it a good testing case. We set the minimum and maximum size of the micro communities to respectively 50 and 100, and to 500 and 1000 for the macro communities. In all networks the average node degree $\bar{d}$ is set to 10 and the maximum degree to 50. For this experiment we use two sets of networks: one with $10^4$ nodes (and $\approx 10^5$ edges) and the other with $10^5$ nodes (and $\approx 10^6$ edges). For each network set we vary the mixing parameters $\mu_1$ and $\mu_2$ setting the fraction of edges between nodes respectfully belonging to different macro and micro communities. We also assume that $\mu_2 > \mu_1$ which means that out of all the edges going out of a micro-community, more edges point towards another micro-community within the same macro-community rather than towards another macro-community. We will use the following $(\mu_1, \mu_2)$ pairs: $(0.1, 0.2), (0.1, 0.4), (0.2, 0.4), (0.4, 0.5)$. For stability optimisation (SO) on scales greater than 1 we use the threshold $\tau = 0.001$. The $A$ values used here respectively for the $n = 10^4/n = 10^5$ networks are 100/1000 for Reichardt and Bornholdt's (RB), $1000/10000-r_{asymp}$ for Arenas et al.'s (AFG), 5/1 for stability optimisation, 0.1/0.1 for Ronhovde and Nussinov's (RN) and 2/2 for Lancichinetti et al.'s (LFK) and Huang et al.'s (HSLSW) (see below about the relevant scale range observations). Each scale range contains 100 values (i.e. $X = 100$). The results are presented in Table 1 and report for each run on each network the number of community sets uncovered and their parameter range. Figure 1 shows the plotted results for the first network with $n = 10^4$, $\mu_1 = 0.1$, and $\mu_2 = 0.2$.

Considering the results of the two network sets, we can observe that the performances are similar for each criteria on a given network edges distribution disregarding of the size of the network. Indeed multiplying the size of the network by 10 did not affect the accuracy performance. We can also observe that by multiplying by 10 the size of the network, the intervals of relevance for the criterion are either about the same (RN, LFK, HSLSW), multiplied by a factor of 10 (RB, AFG) or divided by a factor 10 (SO). This observation yields a first insight into how relevant scales can relate to the network size (here assuming other properties of the network remain unchanged).

Then overall, all criteria perform well on the first network ($\mu_1 = 0.1$ and $\mu_2 = 0.2$) and then performance decreases as the network gains noise by increasing the $\mu$ values. We can also observe that the global criteria versions of the algorithm are more robust to noise than the local criteria versions. This seems to indicate that within the context of this work the tested local criteria are

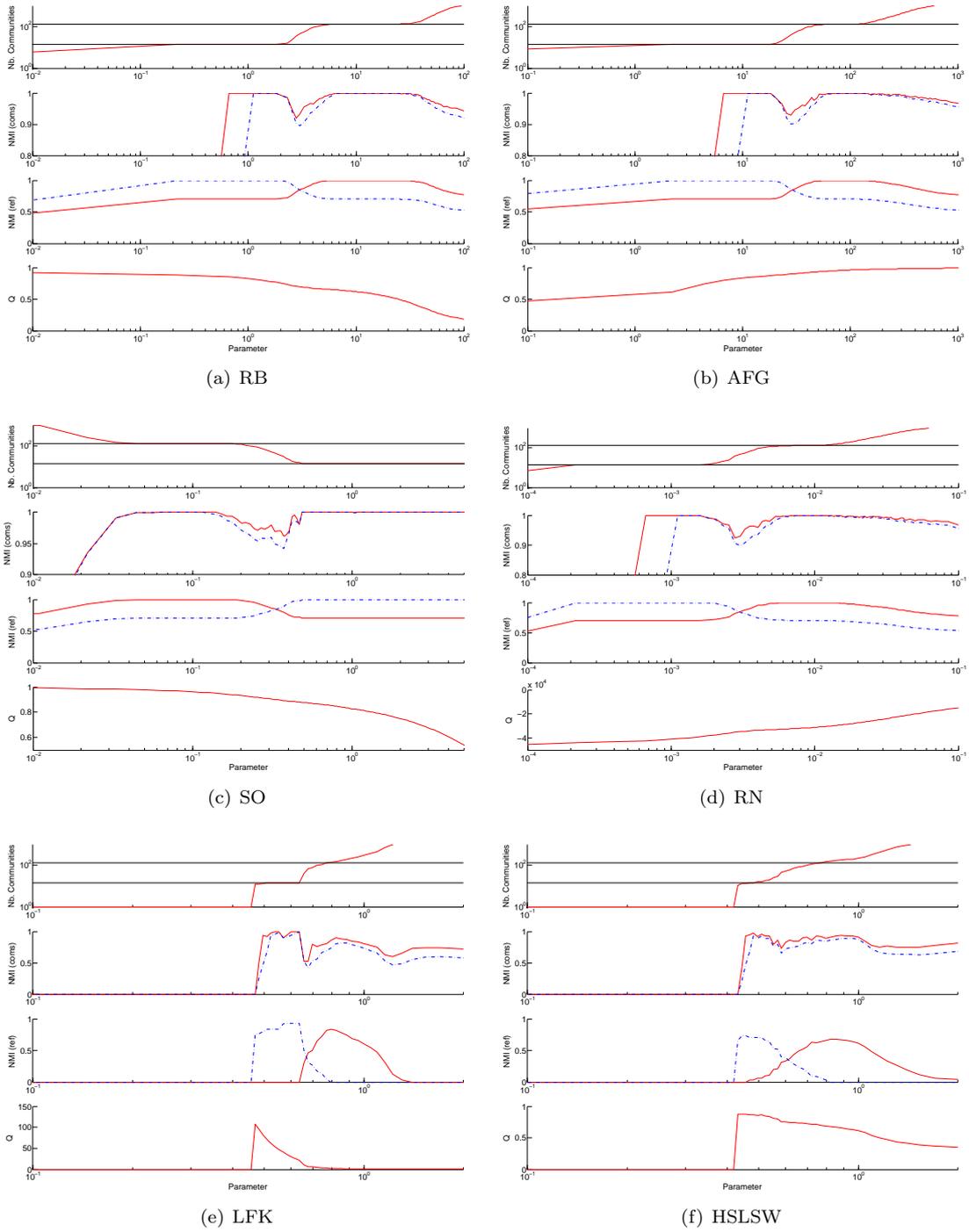

Figure 1: Algorithms' results for the 6 criteria along the scale parameter using a generated network with $10^4$ nodes, about $10^5$ edges and mixing parameters $\mu_1 = 0.1$ and $\mu_2 = 0.2$. The top plot indicates the number of communities uncovered. The two intended community set's size are shown in black straight lines. The second plot shows the averaged NMI between consecutive results: 3 in (red) full and 5 in (blue) dashed. The third plot shows the NMI with the two intended partitions: in (red) full the micro communities and in (blue) dashed the macro communities. The fourth plot shows the Q value corresponding to the returned community sets.

Table 1: Scale parameter range where the macro and then micro communities were spotted. Clearly identified ranges use the interval notation [], values of interest with no clear stable range but a clear NMI peak (weak detection) are given using the notation () and the empty set denotes no detection of the community scale. For SO, the scale parameter works in the opposite way compared to the other criteria. The first network's results are shown in Figure 1.

Networks with $n = 10^4$ and $m \approx 10^5$

| Criteria | $\mu_1 = 0.1$, $\mu_2 = 0.2$ | $\mu_1 = 0.1$, $\mu_2 = 0.4$ | $\mu_1 = 0.2$, $\mu_2 = 0.4$ | $\mu_1 = 0.4$, $\mu_2 = 0.5$ |
|---|---|---|---|---|
| RB | [0.2,2] [5,25] | [0.2,4] [10] | [0.35,2] (20) | [0.55,1] ∅ |
| AFG | [1.5,20] [50,100] | [2,25] [100] | [4,18] (200) | [5,10] ∅ |
| SO | [0.5,5] [0.04, 0.2] | [0.25,5] [0.08] | [0.35,5] (0.1) | [1,5] ∅ |
| RN | [0.0002,0.002] [0.005,0.015] | [0.0002,0.002] [0.015] | [0.00045,0.0015] (0.015) | (0.001) ∅ |
| LFK | [0.45,0.6] [0.76,0.78] | [0.45,0.6] ∅ | ∅ ∅ | ∅ ∅ |
| HSLSW | [0.42,0.5] [0.76,0.78] | [0.55,0.65] ∅ | ∅ ∅ | ∅ ∅ |

Networks with $n = 10^5$ and $m \approx 10^6$

| Criteria | $\mu_1 = 0.1$, $\mu_2 = 0.2$ | $\mu_1 = 0.1$, $\mu_2 = 0.4$ | $\mu_1 = 0.2$, $\mu_2 = 0.4$ | $\mu_1 = 0.4$, $\mu_2 = 0.5$ |
|---|---|---|---|---|
| RB | [2,18] [50,250] | [2,40] [100] | [2,30] (200) | [2,10] ∅ |
| AFG | [15,200] [500,1000] | [20,200] [1500] | [20,100] (2000) | [20,30] ∅ |
| SO | [0.055,1] [0.004, 0.02] | [0.025,1] [0.0065,0.009] | [0.004,1] (0.004,0.007) | [0.25,1] ∅ |
| RN | [0.0002,0.002] [0.005,0.01] | [0.0002,0.002] [0.015] | [0.0002,0.001] (0.02) | [0.0002,0.001] ∅ |
| LFK | [0.45,0.6] [0.77,0.81] | [0.45,0.7] ∅ | [0.6,0.7] ∅ | ∅ ∅ |
| HSLSW | [0.4,0.45] [0.75,0.8] | (0.5,0.65) ∅ | (0.8) ∅ | ∅ ∅ |

less robust to noise than the tested global criteria.

We can also observe that the macro communities are easier to detect than the micro communities. For instance the second and the fourth networks both have communities set with $\mu = 0.4$. In the former, the micro community has a mixing factor of 0.4 ($\mu_2 = 0.4$) whereas in the latter the macro community has a mixing factor of 0.4 ($\mu_1 = 0.4$). We can observe with the first three global criteria (RB, AFG, SO) that the macro communities are well detected even with $\mu_1 = 0.4$ (fourth network) while the micro communities are not (second network). Therefore the community size and not just the leaving edges ratio matters for the detection of communities. The bigger the communities the easier to uncover.

Comparatively, the first three global criteria (RB, AFG, SO) have similar performances. This is perhaps not surprising considering they all derive from the modularity equation with respectively three different ways to manage scales. RN seems to be less robust to noise than the other global criteria (which is consistent with the result of previous experiments on real datasets from [10]). The local criteria both perform well only with little noise $\mu <= 0.2$ and decrease significantly in performance as noise increases. Between the two LFK performs better and copes better with noise. Overall the global criteria seem to be more accurate as their performances decrease less rapidly as noise increases than local criteria's. This study may suggest that the local approach to community detection is less robust to noise than a global approach.

## 4.2 Speed Performance and Memory Usage Testing

In order to evaluate the speed efficiency of our algorithm we use networks generated as in the previous set of experiments and fix the values $\mu_1 = 0.1$ and $\mu_2 = 0.2$. Any configuration could have equally be chosen. We picked one that can be analysed with high accuracy using all criteria.

Several tests are run. The first one measures the overall speed performance and memory usage of the algorithms for each criterion against networks of increasing size up to $10^7$ edges. (Larger networks could be assessed with more RAM.) We vary the number of nodes $n$ and hence the number of edges $m \approx 10n$ in the network. As highlighted above we do not vary the structure (fixed $\mu$ values) as the network structure can also impact the running time of a given algorithm. We also fix the scale ranges as they can have an impact of the running time. We chose $A$ values that reflect intervals of relevance observed in the previous set of experiments: 100 for RB, $1000 - r_{asymp}$ for AFG, 1 for SO starting from 0.01, 0.01 for RN and 1 for LFK and HSLSW. The tests consisted of 10 runs of the algorithm for each criterion on the scale range sampled with 100 values. The results of this first test are given in Figure 2 giving for each value of $m$ the average running time over the 10 runs.

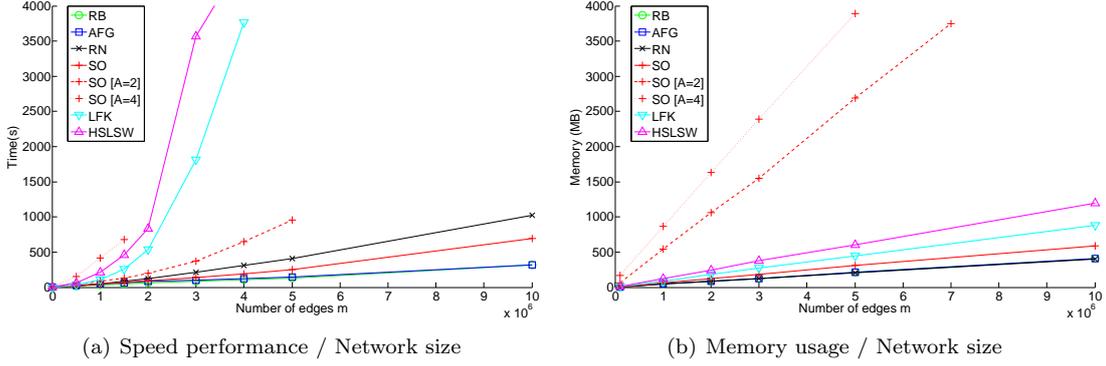

(a) Speed performance / Network size  (b) Memory usage / Network size

Figure 2: Speed performance (a) and memory usage (b) for all criteria given the network's size in edges $m \approx 10n$ up to networks with $m = 10^7$.

We can observe on Figure 2(a) that the running time of the global algorithms with the 4 global criteria (discarding SO with $A > 1$) grows linearly with the number of edges. This confirms the theoretical complexity of the global criteria algorithm in $\mathcal{O}(m)$ and highlights the scalability of the method. In this test $RB$ and $AFG$ are the fastest (curves overlapping). A network with $10^6$ nodes and $10^7$ edges is processed on 100 different scales in only 316 seconds (5 min 26 s) using RB. For SO with $A > 1$ we plot $A = 2$ and $A = 4$. As previously discussed the complexity for $A = 2$ raises to $\mathcal{O}(n \cdot \bar{d}^2) = \mathcal{O}(m \cdot \bar{d})$ hence the greater constant factor compared to $A = 1$. (We evaluated networks up to $m = 5.10^6$ as for a walk of length 2 the resulting network has about $m = 5.10^7$ edges which reaches the memory limit of our machine. Once this limit is reached the use of the swap memory prevents any real-time measurement.) For $A = 4$ the complexity is $\mathcal{O}(n \cdot \bar{d}^4)$ which gets closer to the theoretical limit $\mathcal{O}(n^2 \cdot \bar{d})$ and the networks get very dense. The constant factor is therefore greater than for $A = 2$. (For the same memory limitation reasons the largest network for which the time could be measured accurately has $m = 1.5 \cdot 10^6$ edges.) The local criteria algorithms with the 2 local criteria grow in $\mathcal{O}(n^2)$ which is also consistent with the theoretical complexity analysis.

Figure 2(b) shows that the memory usage remains linear for each criterion with respect to the network size. Each algorithm requires an initial space in $\mathcal{O}(n+m)$ to store the adjacency list and some state variables. The observed results are thus consistent with the theoretical analysis. For the largest network tested above with $m = 10^7$ the RAM usage on the global algorithm is about 400 MB for RB, AFG and RN (curves are overlapping), about 590 MB for SO (due to the storage of the current walk network which takes $\mathcal{O}(m)$ additional space). For SO, the cost increases as the random walk lengthens, as can be seen from the curves for $A = 2$ and $A = 4$.

For the local algorithm, nodes can belong to several communities hence the greater amount of space needed of about 880 MB for LFK and about 1200 MB for HSLSW (due to the additional similarity values between nodes taking $\mathcal{O}(m)$ space).

Larger networks (i.e. $m \geq 10^8$) could be tested with more RAM on the machine. The linear curves from Figure 2 clearly show the evolution in time and space requirement as the network size grows. For example we can work out that about 4 GB of free RAM would be required to store a network with $m = 10^8$ using RB, AFG or RN considering that about 400 MB are needed for a network with $m = 10^7$.

The second test measures the running time of all algorithms, averaged over 10 runs, on a network of size $n = 10^4$ and $m = 10^5$ generated as described above but varying the number of values in the scale range (i.e. $X$). This shows the impact of the scale range sampling on the efficiency. The results are given in Figure 3. Note that the performance of a run also depends on the scale range boundaries as the work performed by the algorithm depends on the criterion, which varies along the scale. The same algorithm on the same network with the same amount of scale values can be faster with one range of values than with another. Therefore the fact that one algorithm runs a bit faster than another one in this context does not mean that it would always run faster on any scale range.

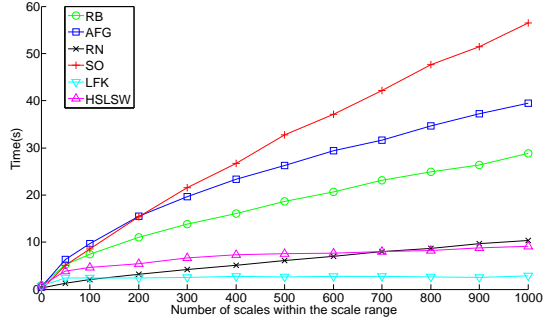

Figure 3: Speed performance for all criteria given the number of scale values in the scale range.

The results from Figure 3 show that the cost of refining the scale range for all algorithms grows sub-linear with respect to the number of scales, and remains almost constant for LFK and HSLSW. The principle of our method is to uncover communities at a scale given the result at the previous scale. This result confirms that the overhead of an additional scale is greatly reduced. For SO, at each scale value the weights have to be recomputed for the network reflecting the current random walk, hence the overhead observed on its curve compared to the curve of the other global criteria.

## 5 Conclusion

In this paper we presented a fast method for the detection of communities in networks across scales. The algorithm builds at each scale upon the results found at the previous scale and only computes the changes brought by the scale variation. The processing is done in two phases for each scale parameter value. First subtle changes are performed at the node level, then coarser changes are performed at the community level. The method is criteria independent and has been derived into two algorithms: the first one for global criteria, the second one for local criteria. We implemented 6 known criteria, 4 global and 2 local, taken from the relevant literature and developed heuristics for efficient implementations. The complexity for the global criteria algorithm is $\mathcal{O}(m)$ with crisp community boundaries. The local criteria algorithm has a complexity of $\mathcal{O}(n^2)$ but allows communities to overlap. Experiments have demonstrated the speed efficiency and the accuracy of the algorithms with respect to each criterion. Networks of up to $10^6$ nodes and $10^7$ edges were analysed and the limitation in network size was due to our memory limits of 4GB. We used a regular desktop machine for experiments therefore demonstrating the potential and performance an average user can expect using our method. On our machine a network with $10^7$ edges is processed accurately over 100 scales in about 5 minutes.

This paper also made a comparative analysis of the 6 considered multi-scale criteria, within the scope of this algorithm. Our study revealed that the global criteria seem to be more robust to noise and thus more accurate than local criteria. Indeed experiments showed that the performances of the former decrease less rapidly as noise increases than the latter's. Also global criteria algorithms are more speed efficient. Yet, assuming an appropriate community merging phase, only local approaches could deal with huge networks that are too large to fit in memory. Such approaches are also better suited to distributed structures such as the internet.

Considering that user analysis is often an important part in data analysis, future work could consider adding a visualisation framework to offer interactive possibilities such as visually exploring the community sets and refining some parts of the scale range for further analysis. Also the algorithms could be parallelised. Whether node shifting or community growth, these tasks are fairly independent and could benefit from some parallel computation. Similarly the merging phase could handle in parallel several subsets of communities. Another optimisation could consider collapsing nodes when they form a solid core of a community. This could reduce the complexity as the algorithm progresses.

In a different direction, future work could also consider networks changing over time and use the two phases to track the evolution of communities. As the process would no longer be aggregative, additional operations such as community splitting would be required.

Finally the framework developed for this work is freely available for download [3]. It has been designed to easily integrate additional algorithms in order to offer a platform usable for data analysis and algorithm comparison.

**Acknowledgements:** This work was conducted as part of the *Making Sense*[4] project financially supported by EPSRC (project number EP/H023135/1).

---

[3] http://www.elemartelot.org
[4] http://www.making-sense.org